\documentclass[manuscript]{emulateapj}
\newcommand{\mum}{${\rm \mu m}$} 
 \newcommand{\ergs}{${\rm erg~s^{-1}}$} 
 \newcommand{\msun}{M$_\odot$} 
  
\newcommand{\kms}{\,\hbox{km}\,\hbox{s}^{-1}}
 
 \newcommand{\lx}{$L_{\rm X}$} 
 \newcommand{\lir}{$L_{\rm IR}$}

\def\nar{NewAR} 
\def\simgt{\lower.5ex\hbox{\gtsima}} 
\def\simlt{\lower.5ex\hbox{\ltsima}} 
 
\shorttitle{NO SIGNATURE OF SUPPRESSED STAR FORMATION AMONG LUMINOUS
AGNS} 
\shortauthors{HARRISON ET AL.} 
 
\begin{document} 
 
\title{No clear submillimeter signature of suppressed star formation among
  X-ray luminous AGNs} 
 
\author{
C.\ M.\ Harrison\altaffilmark{1}
D.\ M.\ Alexander\altaffilmark{1},
J.\ R.\ Mullaney\altaffilmark{1},
B.\ Altieri\altaffilmark{2},
D.\ Coia\altaffilmark{2},
V.\ Charmandaris\altaffilmark{3},
E.\ Daddi\altaffilmark{4}, 
H.\ Dannerbauer\altaffilmark{5}, 
K.\ Dasyra\altaffilmark{6},
A.\ Del~Moro\altaffilmark{1},
M.\ Dickinson\altaffilmark{7},  
R.\ C.\ Hickox\altaffilmark{8}, 
R.\ J.\ Ivison\altaffilmark{9},
J.\ Kartaltepe\altaffilmark{7},
E.\ Le Floc'h\altaffilmark{4},
R.\ Leiton\altaffilmark{4,10},
B.\ Magnelli\altaffilmark{11}, 
P.\ Popesso\altaffilmark{11},
E.\ Rovilos\altaffilmark{1},
D.\ Rosario\altaffilmark{11},
A.\ M.\ Swinbank\altaffilmark{1}
}
\altaffiltext{1}{Department of Physics, Durham University, South Road,
  Durham DH1 3LE, U.K.}
\altaffiltext{2}{Herschel Science Centre, European Space Astronomy Centre, Villanueva de la Canada, 28691 Madrid, Spain}
\altaffiltext{3}{Department of Physics \& Institute of Theoretical and Computation, Physics, University of Crete, 71003 Heraklion, Greece}
\altaffiltext{4}{Laboratoire AIM, CEA/DSM-CNRS-Universit\'{e} Paris Diderot, Irfu/Service d Astrophysique, CEA-Saclay, Orme des Merisiers, 91191 Gif-sur-Yvette Cedex, France}
\altaffiltext{5}{Universit\"at Wien, Insitut f\"ur Astrophysik, T\"urkenschanzstra\ss e 17, A-1180 Wien, Austria}
\altaffiltext{6}{Observatoire de Paris, LERMA (CNRS:UMR8112), 61 Av. de l' Observatoire, F-75014, Paris, France}
\altaffiltext{7}{National Optical Astronomy Observatory, 950 North Cherry Avenue, Tucson, AZ 85719, USA}
\altaffiltext{8}{Department of Physics and Astronomy, Dartmouth College, 6127 Wilder Laboratory, Hanover, NH 03755, USA}
\altaffiltext{9}{UK Astronomy Technology Centre, Royal Observatory,
  Blackford Hill, Edinburgh EH9 3HJ, UK}
\altaffiltext{10}{Department of Astronomy, Universidad de Concepci\'on, Casilla 160-C, Concepci\'on, Chile} 
\altaffiltext{11}{Max-Planck-Institut f\"ur Extraterrestrische Physik (MPE), Postfach 1312, 85741, Garching, Germany}

\begin{abstract} 
 
  Many theoretical models require powerful active galactic nuclei 
  (AGNs) to suppress star formation in distant galaxies and reproduce 
  the observed properties of today's massive galaxies. A recent study 
  based on {\it Herschel}-SPIRE submillimeter observations claimed to provide 
  direct support for this picture, reporting a significant decrease in 
  the mean star-formation rates (SFRs) of the most luminous AGNs 
  (\lx$>$$10^{44}$\,\ergs) at $z$\,$\approx$\,1--3 in the {\it Chandra} Deep 
  Field-North (CDF-N). In this Letter we extend these results using 
  {\it Herschel}-SPIRE 250~\mum\ data in the COSMOS and CDF-S fields to achieve 
  an order of magnitude improvement in the number of sources at 
  \lx$>$$10^{44}$\,\ergs. On the basis of our analysis, we find no strong evidence for suppressed star formation in 
  \lx$>$$10^{44}$\,\ergs\ AGNs at $z$\,$\approx$\,1--3. The mean SFRs of the 
  AGNs are constant over the broad X-ray luminosity range of \lx$\approx$$10^{43}$--$10^{45}$\,\ergs\ (with 
  mean SFRs consistent with typical star-forming galaxies at
  $z$\,$\approx$\,2; $\langle{\rm
    SFRs}\rangle$$\approx$100--200\,\msun\,yr$^{-1}$). We suggest that
  the previous CDF-N results were likely due to low number statistics. We discuss our results in the context of current 
  theoretical models. 
 
\end{abstract} 
 
 
\section{Introduction} 
 
Over the last decade it has become increasingly clear that active 
galactic nuclei (AGNs) have had a profound influence on the formation 
and evolution of galaxies. Some of the most compelling evidence is 
based on computational simulations and theoretical models, which have 
shown that AGNs can suppress or shut down star formation by heating, or 
removing, the cold gas in their host galaxies (e.g., \citealt{Silk98, 
  Springel05, DiMatteo05, Bower06, Hopkins06, Hopkins08a, 
  DeBuhr12}). While there is some indirect observational support for the 
suppression of star formation by AGN activity (see 
\citealt{Alexander12} and \citealt{Fabian12} for general reviews), we 
currently lack conclusive observational evidence across the overall
population.
 
One approach to assess the influence of AGNs on the growth of galaxies 
is to explore possible connections between AGN activity and star 
formation. A large suite of studies have investigated the relationship 
between AGN activity and star formation out to high redshifts
($z \approx 3$), using deep X-ray and infrared data to constrain the AGN activity and star 
formation rates (SFRs), respectively (\citealt{Mullaney10, Hatziminaoglou10, Lutz10, Shao10, Mullaney12a, Santini12, Rosario12, 
  Page12, Rovilos12}). These studies have shown that the mean SFRs of 
moderate-luminosity AGNs (\lx$\approx10^{42}$--$10^{44}$\,\ergs) out to 
$z\approx$~3 are comparable to those of co-eval star-forming galaxies 
(e.g.,\ \citealt{Noeske07, Elbaz07, Daddi07, Pannella09}). However, while these 
studies find broadly similar results for the average SFRs of 
moderate-luminosity AGNs, the picture for luminous AGNs 
(\lx$>10^{44}$\,\ergs) is less clear, with the majority of these 
studies arguing that the mean SFR either rises or remains flat towards 
the highest luminosities (e.g.,\ \citealt{Lutz10, Shao10, Rosario12, 
  Rovilos12}). 
 
Recently, \cite{Page12} used {\it Chandra} X-ray and {\it 
  Herschel}-SPIRE submillimeter data in the {\it Chandra} Deep Field-North 
(CDF-N) to report that the mean SFRs of luminous AGNs 
(\lx$>10^{44}$\,\ergs) at $z\approx$~1--3 are significantly lower than 
those of moderate-luminosity AGNs. The implications of this study are 
potentially very significant as they imply a direct, empirical connection 
between luminous AGN activity and the suppression of star 
formation. However, these results are in disagreement 
with other studies, many of which also used {\it Herschel} data in 
similarly deep fields (e.g., \citealt{Lutz10, Shao10, Rosario12, 
  Rovilos12}). 

The aim of this Letter is to extend the \cite{Page12} 
study using {\it Herschel}-SPIRE 250\,\mum\ data in the wider-area 
COSMOS field, improving the source statistics for 
\lx$>10^{44}$\,\ergs\ AGNs by an order of magnitude. Indeed, a clear 
limitation of the \cite{Page12} study was the small number of sources 
in the \lx$>10^{44}$\,\ergs\ bins ($\approx$~7--14 objects). We
  note that our Letter only explores the mean SFRs using stacking
  analysis and does not explore the individual detection rates (i.e.,
  Figure~1 of \citealt{Page12}); however, significant differences in the
  detection fractions should lead to differences in the mean SFRs. We also 
repeat our analysis using {\it Herschel}-SPIRE 250\,\mum\ data in the CDF-N and {\it Chandra} Deep Field South (CDF-S) 
fields to explore the effect of using smaller fields. In our 
analysis we use $H_{0}=71\kms$, $\Omega_{\rm{M}}=0.27$, 
$\Omega_{\Lambda}=0.73$ and assume a Salpeter initial mass function 
(IMF). 
 
%%%%%%%%%%%%%%%%%%%%%%%%%%%%%%%%%%%%%%%%%%%%%%%%%%%%%%%%%%%%%%%%%%%%%%%%%%%%% 
 
%\newpage 
 
\section{Catalogs and data} 
\label{Sec:Data} 
 
\subsection{X-ray data} 
 
\begin{figure} 
\includegraphics[angle=90, scale=0.4]{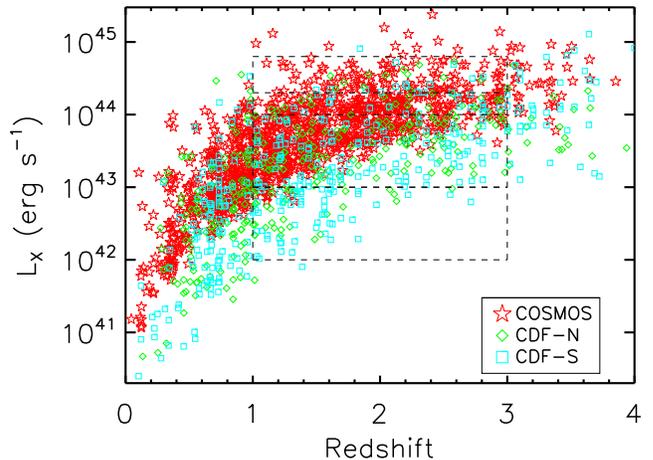} 
\caption{Rest-frame 2--8\,keV luminosities (not-absorption corrected)
  of the X-ray AGNs detected in the three deep fields (see
  \S\ref{Sec:Data} for details) versus redshift. The areas of parameter space used in our
  stacking procedures (see Figure~\ref{fig:final}) are illustrated with the dashed lines. Caution
  should be taken when interpreting the lowest X-ray luminosity bin ($L_{\rm X}$=10$^{42}$--10$^{43}$\,erg\,s$^{-1}$) as
it is incomplete at $z$$\gtrsim$2.} 
\label{fig:LvZ} 
\end{figure}

Our samples of AGNs are X-ray selected and cover a broad range of 
luminosities (\lx$=10^{42}-10^{45}$\,\ergs) over the redshift range 
$z=1-3$ (see Figure~\ref{fig:LvZ}). We use the CDF-N (\citealt{Alexander03}), CDF-S (\citealt{Xue11}) and 
COSMOS (\citealt{Elvis09}) AGN samples to obtain statistically 
meaningful numbers of AGNs in key \lx\ ranges. Redshifts for the CDF-N
and CDF-S samples were taken from Bauer et~al. in prep. and \cite{Xue11}, respectively. Redshifts for the 
COSMOS sample were taken from \cite{Civano12}. Unless otherwise 
stated, we include both photometric and spectroscopic redshifts, which 
comprise $\approx53\%$ and $\approx47\%$ of the combined samples 
respectively; however, we note that our conclusions do not 
change if we only consider spectroscopic redshifts. Following 
\cite{Page12} we derive the rest-frame observed (i.e., not 
absorption-corrected) 2--8~keV luminosity ($L_{\rm X}$) of AGNs in
each field using $L_{\rm X}=4 \pi D^2F_{\rm X}(1+z)^{(\Gamma-2)}$, where $F_{\rm X}$ is 
the observed X-ray flux\footnote{We note \cite{Elvis09} report 2--10~keV fluxes, which we convert to 2--8~keV 
  fluxes by assuming $\Gamma=1.9$ (factor of $\approx$0.85 correction).} (2--8~keV), $D$ is the luminosity 
distance, $z$ is the redshift and $\Gamma$ is the spectral index used for $k$-corrections (we assume $\Gamma=1.9$).

In Figure~\ref{fig:LvZ} we show $L_{\rm{X}}$ versus redshift for the X-ray AGNs in each of these fields.  These three samples 
are highly complimentary; the ultra-deep CDF-N and CDF-S samples 
provide large numbers of moderate luminosity AGNs at $z=1-3$ (i.e., 
\lx$=10^{42}-10^{44}$\,\ergs) while the shallower, wider-area COSMOS 
survey populates the high-luminosity ($L_{\rm X}>10^{44}$\,\ergs) 
portion of the parameter space largely missed by these small area
surveys. 
 
\begin{figure} 
\includegraphics[angle=90, scale=0.4]{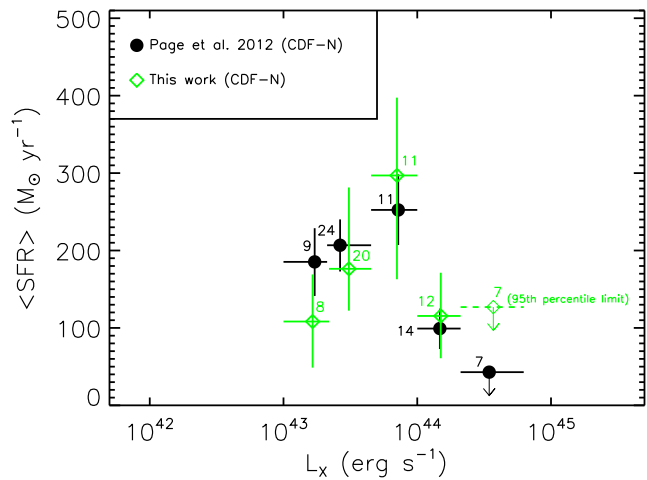} 
\caption{Mean star formation rate (SFR) vs. X-ray luminosity for
  AGNs in the CDF-N field compared to \cite{Page12}; see
  \S\ref{Measuring} for details. We find excellent agreement between our mean SFRs and
  those of \protect\cite{Page12} across all \lx\ bins; we indicate the 
  number of sources in each bin (the differences in the number of 
  objects between our study and \cite{Page12} are due to 
  differences in the redshift catalogs). We note that for
    \cite{Page12} the values were extracted directly
    from their figures as tabulated values were unavailable. As with 
  \protect\cite{Page12}, we do not measure a significant flux in the highest \lx\ bin for which we also include the 95th 
  percentile limit (see \S\ref{Measuring}). This limit is broadly consistent with the mean 
  SFRs for our lower \lx\ bins.} 
\label{fig1} 
\end{figure} 

\begin{figure*} 
\includegraphics[angle=90, scale=0.8]{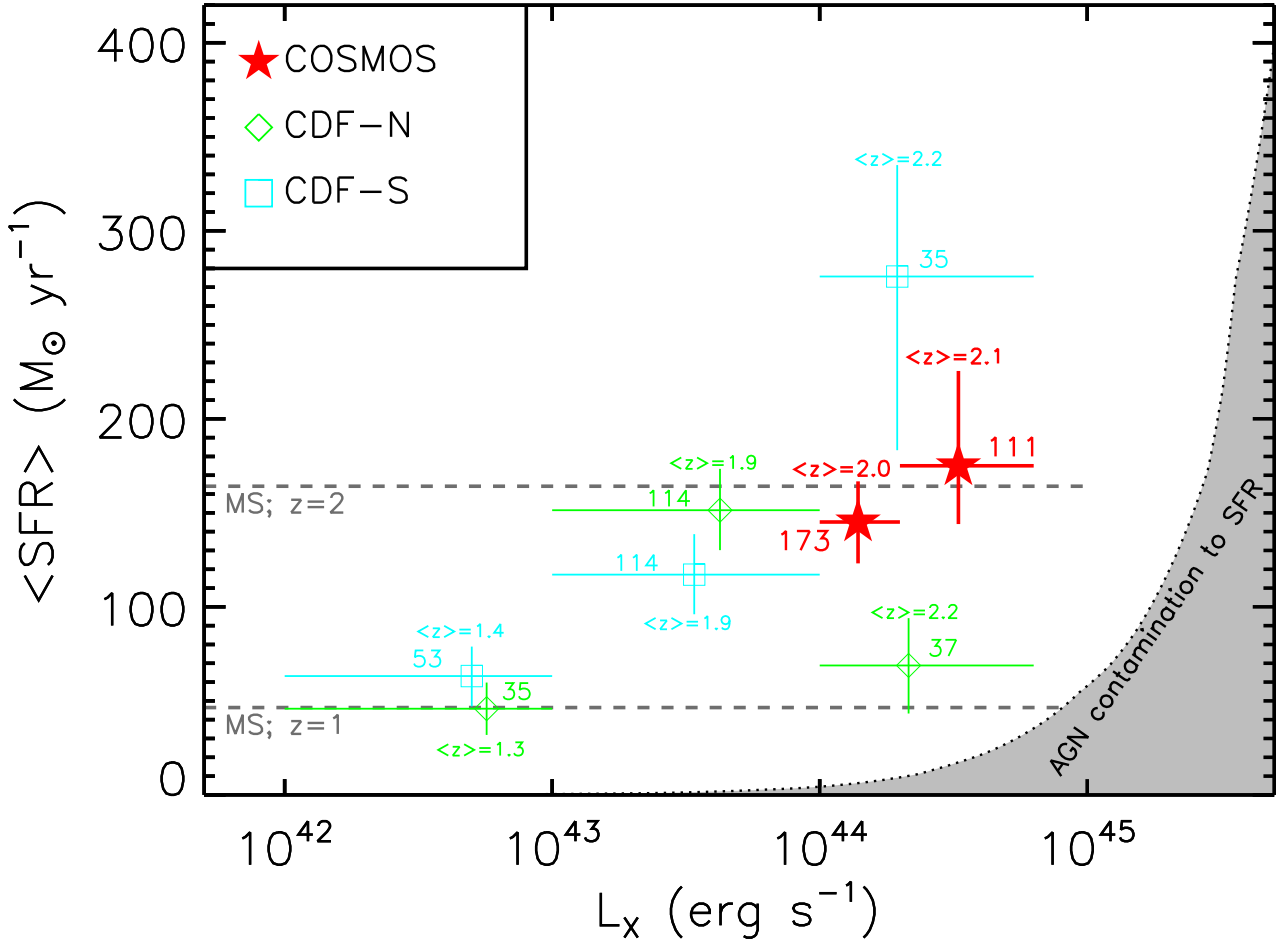} 
\caption{Mean SFR vs. $L_{\rm{X}}$ for AGNs in the three fields (see \S\ref{Measuring} for details). We indicate the 
  number of, and the average redshift of, the AGNs in each bin (Table~1). The dotted curve indicates the contamination to the measured SFR from AGN 
  activity (i.e., the amount that the SFRs will be boosted due to
  contamination of the 250\,\mum\ flux [rest-frame 83\,$\mu$m at 
  $z=2$] by AGN activity) and is calculated using the \lx-$L_{\rm MIR}$ relation of 
  \protect\cite{Gandhi09} and the intrinsic AGN SED of 
  \protect\cite{Mullaney11}. The dashed lines show the average SFR
  expected (derived from \citealt{Elbaz11}) for a galaxy with a stellar mass of $9\times10^{10}$\,\msun\ galaxy, (i.e., about the mean stellar 
  mass of X-ray detected AGN hosts; e.g., \citealt{Xue10}; \citealt{Mullaney12a}) at $z=1.0$ 
  and $z=2.0$. We find that the mean SFR distribution at 
  \lx$>10^{42}$\,\ergs\ is roughly equal to that expected from typical 
  star-forming galaxies at the average redshifts of our bins (our
  \lx$<10^{43}$\,\ergs\ bins have lower mean redshifts and hence lower
  mean SFRs; e.g., \citealt{Mullaney12a}). We show that there
  are variations in the measured mean SFRs between the small CDF-N and
  CDF-S fields and argue that this, combined with low
  number statistics drive the \cite{Page12} results shown in Figure~\ref{fig1}.} 
\label{fig:final} 
\end{figure*}

\begin{table} 
\begin{center} 
  \caption{Derived Average Properties of Sub-samples}\label{table1} 
\begin{tabular}{@{}lcccccc@{}} 
\hline 
\hline 
$L_{\rm X}$ range           &      $N_{\rm AGN}$&$\langle{z}\rangle$   &$\langle{L_{\rm X}}\rangle$&$\langle{S_{250}}\rangle$&$\langle{\rm SFR}\rangle$\\ 
(1)                                &(2)                &(3)                                      &(4)                                    &(5)                                    &(6)                                     \\ 
\hline 
\multicolumn{6}{l}{CDF-N  (HerMES; Figure 2)*} \vspace{0.1cm}\\ 
$43.00-43.33$            &                 8&1.8                              &$43.2$&$5.4^{+3.0}_{-3.0}$ 	                         &$108^{+61}_{-60}$                     \\ 
$43.33-43.66$            &               20&1.7                             &$43.5 $&$9.4^{+2.3}_{-2.4}$                              &$176^{+105}_{-54}$                     \\ 
$43.66-44.00$            &               11&1.9                              &$43.8 $&$10.9^{+3.7}_{-3.6}$                           &$297^{+100}_{-134}$                     \\ 
$44.00-44.33$            &               12&2.0                              &$44.2 $ &$4.8^{+2.5}_{-2.4}$                              &$115^{+56}_{-55}$                     \\ 
$44.33-44.80$            &                 7&2.1                              &$44.6  $&$<$4.4                      &$<127$                     \\ 
&&&&\\ 
\multicolumn{6}{l}{CDF-N  (GOODS-{\it H}; Figure 3)} \vspace{0.1cm}\\ 
$42.00-43.00$            &               35&1.3                            &$42.8 $&$3.5^{+1.1}_{-1.1}$                              &$46^{+14}_{-14}$                     \\ 
$43.00-44.00$            &               114&1.9                          &$43.6 $&$6.8^{+1.0}_{-0.9}$                              &$151^{+22}_{-21}$                     \\ 
$44.00-44.80$            &               37&2.2                            &$44.3 $&$2.2^{+0.9}_{-0.9}$                             &$69^{+25}_{-26}$                     \\ 
&&&&\\ 
\multicolumn{6}{l}{CDF-S  (Figure 3)} \vspace{0.1cm}\\ 
$42.00-43.00$            &               53&1.4                            &$42.7 $&$4.3^{+1.0}_{-1.1}$                              &$63^{+16}_{-16}$                     \\ 
$43.00-44.00$            &               114&1.9                          &$43.5 $&$5.3^{+1.0}_{-1.0}$                              &$117^{+22}_{-21}$                     \\ 
$44.00-44.80$            &               35&2.2                            &$44.3  $&$7.2^{+1.4}_{-1.4}$                             &$276^{+59}_{-92}$                     \\ 
&&&&\\ 
\multicolumn{6}{l}{COSMOS (Figure 3)} \vspace{0.1cm}\\ 
$44.00-44.30$            &               173&2.0                            &$44.1  $&$6.0^{+0.9}_{-0.9}$                             &$145^{+22}_{-22}$                     \\ 
$44.30-44.80$            &               111&2.1                            &$44.5  $&$6.3^{+1.1}_{-1.1}$                             &$175^{+50}_{-31}$                     \\ 
\hline 
\end{tabular} 
\end{center} 
\tablecomments{(1) X-ray luminosity (2--8\,keV) range of each 
  bin ($\log$[erg\,s$^{-1}$]); (2) Number of X-ray AGNs in each bin;
  (3) Mean redshift; (4) Mean \lx\ ($\log$[erg\,s$^{-1}$]); (5) Mean
  250\,\mum\ flux and uncertainties (mJy); (6) Mean star formation
  rate (\msun\,yr$^{-1}$). The quoted uncertainties are derived from
  our bootstrap method (see \S\ref{Measuring}). *To be consistent with
  \cite{Page12} we use the same $L_{\rm{X}}$ bins and only include AGNs with spectroscopic redshifts when 
  stacking the HerMES data for the CDF-N field (see \S\ref{Results}).} 
\end{table} 

\subsection{IR Data} 
 
The mean SFRs for our AGN samples are derived from 
250\,\mum\ observations taken by the SPIRE instrument on board {\em 
  Herschel} which are relatively unaffected by contamination from the AGN (e.g., 
\citealt{Hatziminaoglou10}) or obscuration due to dust. Our results are primarily based on the observations of 
the CDF-N, CDF-S and COSMOS fields undertaken as part of the HerMES 
campaign (P.I.: S. Oliver; described in \citealt{Oliver12}). We 
downloaded the individual level-2 data products from each scan of 
each field (totaling $\sim$\,14\,hours, $\sim$\,20\,hours and 
$\sim$\,50\,hours in CDF-N, CDF-S and COSMOS, respectively) from the 
{\em Herschel} ESA archive and then aligned and coadded the 
images. The coadded images were then aligned onto a common world 
co-ordinate system. We also checked for consistency 
between our own coadds and those produced using the standard HIPE v9.0 
pipeline. 
 
We also make use of 250\,\mum\ SPIRE observations of the CDF-N field 
that were undertaken as part of the GOODS-{\em Herschel} program 
(GOODS-{\em H}; P.I.: D. Elbaz) and cover the whole of the CDF-N (totaling 44 hours 
when combined with the HerMES observations). The observations and steps used to produce the science images are described in \cite{Elbaz11}. We note that our 
results derived from the GOODS-{\em H} CDF-N observations are 
consistent with those derived from the HerMES observations of the same 
field.
 
%%%%%%%%%%%%%%%%%%%%%%%%%%%%%%%%%%%%%%%%%%%%%%%%%%%%%%%%%%%%%%%%%%%%%%%%%%%%% 
 
\section{Measuring average star formation rates} 
\label{Measuring} 
 
To determine the mean SFRs of AGNs as a function of \lx\ we split our samples into a number of \lx\ bins 
(see Table \ref{table1}).  The majority of the X-ray AGNs are not individually detected by {\em 
  Herschel}-SPIRE (e.g., \citealt{Hatziminaoglou10, Page12}) and
therefore, we rely on stacking analysis to obtain the mean 250\,\mum\ fluxes of our AGN 
samples in each bin. We stack at the optical positions of all X-ray AGNs in our 
samples, whether they are individually 250\,\mum-detected or not
(following \citealt{Page12}). The point-spread function (PSF) of the SPIRE 250\,\mum\ science images is normalized to 1 and in 
units of mJy beam$^{-1}$, such that the mean 250\,\mum\ flux was taken
to be the central pixel value of each stacked image. 
 
To determine if the measured fluxes from our stacked images are
significantly above the background we employed a Monte-Carlo approach, which quantifies the level of 
background and confusion noise (i.e., flux contribution to our stacked images 
from nearby, unassociated sources). First, along with each ``true'' stacked image we also made 10,000 
``random'' stacks, stacking the same number of random positions around 
the field as in our true stacks. The mean flux of the random stacks 
was subtracted from that of the true stack, effectively removing 
any systematic contribution from the background or confusion. Second, we compare our measured ``corrected'' fluxes of our true stacks to the distribution 
of fluxes from our random stacks. We consider a stacked signal to be 
significant if the flux is greater than the flux of 95\%\ of the 
random trials. Following this criterion, all but one of our stacked
images have flux measurements that are significant (see \S\ref{Results} for 
details). We note that this method of randomly choosing positions
to estimate the background does not take into account any clustering between X-ray AGNs and IR-sources (this was also not considered in \citealt{Page12}). While this will
not affect the SFR-\lx\ trends, the mean background
subtracted from each stacked image may be slightly under-estimated,
resulting in the absolute mean SFRs quoted being slightly
over-estimated. However, we note that our results are consistent with
studies that are less affected by confusion noise (e.g.,
\citealt{Mullaney12a}; \citealt{Rosario12}), giving confidence in
our procedures. 

We used the spectral energy distribution (SED) library of \cite{Chary01} to convert mean 250\,\mum\ fluxes from our 
stacked images to mean integrated 8--1000\,\mum\ infrared luminosities 
(\lir), selecting a redshifted \cite{Chary01} SED on the basis of the 
monochromatic luminosity probed by the 250\,\mum\ waveband. These mean \lir\ values were converted to mean 
SFRs using \cite{Kennicutt98}. In the redshift range $z$=1--3 investigated here, the observed 250\,\mum\ fluxes
correspond to the peak of the SEDs in the rest frame. As such
the $L_{250}$ to $L_{\rm{IR}}$ correction factors do not vary much as a function of SED shape
when compared to using shorter wavelengths (see Figure~3 in
\citealt{Elbaz10}). Indeed, when we compared {\em all} of the
  \cite{Chary01} SEDs, redshifted to $z$\,=2, the conversion factors
  were consistent within $\approx$20\%; we also repeated our analysis
using the main-sequence SED from \cite{Elbaz11} for all of the bins and found no significant
difference.  While the choice of SED adds a small additional
uncertainty in the measured SFRs (such that quoted absolute SFRs
  should be used with care), all of the observed trends, and hence the main
conclusions of this Letter, remain unaffected. We also show in
\S\ref{Results} that we reproduce the results of \cite{Page12},
providing further confidence in our approach.
 
Upper and lower limits on the mean SFRs for each of our \lx\ bins were
calculated using a bootstrapping technique, therefore taking into account the distribution of SFRs in each stack. We 
randomly subsample (with replacement) the AGNs in each of our bins,
restack and recalculate the mean SFR. We did this 10,000 times for each bin to 
produce a distribution of mean SFRs. The quoted upper and lower
limits on the mean SFRs correspond to 
the 16th and 84th percentiles (i.e., incorporating 68\%, or 
$\approx\pm1\sigma$) of this distribution (Table~1).
 
\section{Results and discussion} 
\label{Results} 
 
We first stacked sub-samples of X-ray AGNs in the CDF-N. In an 
attempt to reproduce the results of \cite{Page12} we split the 
sample into the same luminosity bins used in their study
(Table~\ref{table1}) and only included AGNs with spectroscopic 
redshifts; see Figure~\ref{fig1}. We find excellent 
agreement between our mean SFRs and 
those of \cite{Page12} (i.e., data points are consistent within
their uncertainties), demonstrating the compatibility of the 
procedures used to derive these results.\footnote{We note that 
  \cite{Page12} used a different technique to that adopted by us here; 
  they derive average SFRs by fitting a modified black body to the SPIRE fluxes of each source.} In agreement with \cite{Page12}, we do not obtain
a significant flux measurement from the stacked 250\,\mum\ image of our highest \lx\ bin 
(log$[$\lx/\ergs$]=44.33-44.80$). However, unlike \cite{Page12}, we find our upper limit is consistent with the mean SFRs of the lowest \lx\ bins.

The non-detection at FIR wavelengths of the AGNs in our highest \lx\
bin in the CDF-N field, combined with the small numbers of AGNs in
this bin (7 sources), clearly demonstrates the need for larger
numbers of sources at high X-ray luminosities.  To achieve this we
performed the same stacking procedures using the COSMOS
survey. The inclusion of the COSMOS AGNs means that the number of
sources in our highest \lx\ bins are now greater by an order of
magnitude (Figure~\ref{fig:LvZ} and Table \ref{table1}). As such, we
can now place tighter constraints on the mean SFRs of
high-luminosity AGNs. On the basis of our stacking analysis in the
COSMOS field, we find that the mean SFRs of AGNs with
\lx$>10^{44}$\,\ergs\ are consistent with those of
\lx=$10^{43}$--$10^{44}$\,\ergs\ AGNs
($\approx$100--200\,\msun\,yr$^{-1}$; see Figure~\ref{fig:final} and
Table~1). This implies that the mean SFR of $z$=1--3 AGNs is
independent of X-ray luminosity (although increases with increasing
redshifts), in broad agreement with \cite{Rosario12} and indicates
that the flat SFR distribution at \lx$<10^{44}$\,\ergs\ (e.g.,
\citealt{Lutz10, Shao10, Mullaney12a, Rosario12}) continues out to at
least \lx$\approx10^{44.8}$\,\ergs.\footnote{We note that this
  approach is different to averaging over {\em all}
$L_{\rm{X}}$ and $L_{\rm{IR}}$ over {\em all} star-forming galaxies (whether X-ray detected or not) where a clear relationship
  between $L_{\rm{X}}$ and SFR is found (see \citealt{Mullaney12b}).}
 
The results derived from our stacking of AGNs in the COSMOS field 
may suggest that poor source statistics and potentially field--to--field 
variations (i.e., cosmic variance) are at least partially responsible for the disagreement 
between the results from COSMOS and CDF-N at high X-ray 
luminosities. To further investigate this we also stacked the SPIRE data in 
the CDF-S field, which is of a similar size as the CDF-N and contains 
approximately the same number of high \lx\ AGNs (Figure~\ref{fig:LvZ}). We also restack the CDF-N AGNs using the deeper 250~\mum\ observations taken as part of the 
  GOODS-{\em H} program since this is of comparable depth to the HerMES 
  CDF-S observations. As shown in Figure~\ref{fig:final}, we find broad agreement between the mean 
SFRs of moderate luminosity AGNs (i.e., $L_{\rm X}<10^{44}$\,\ergs) in 
both fields: the mean SFRs with X-ray luminosity over the range 
$L_{X}\approx10^{42}$--$10^{44}$\,\ergs\ are consistent with those
expected from typical star-forming galaxies with the average redshifts
and typical stellar masses observed for the AGN host galaxies. The increasing
specific SFRs with redshift (e.g., \citealt{Elbaz11}) is likely to
drive the apparent increase in the mean SFRs between $L_{X}\approx10^{42}$--$10^{43}$\,\ergs\
and $L_{X}\approx10^{43}$--$10^{44}$\,\ergs (i.e., they have different
mean redshifts). 

In contrast to the lower X-ray luminosity systems, the mean SFRs of AGNs
between CDF-N and CDF-S appears to diverge at the highest X-ray
luminosities (i.e.,\ $L_{X}>10^{44}$\,\ergs). We note that the
higher mean SFR observed for high \lx\ AGNs in the CDF-S is broadly consistent with the results found by 
\cite{Lutz10} and \cite{Rovilos12} in the same field. Despite
these field--to--field variations the mean SFRs in the CDF-S and CDF-N 
fields are consistent with that found in the COSMOS field (within $\approx$1$\sigma$ and
$\approx$3$\sigma$ respectively).

We now consider if our main result indicates that luminous AGN activity does not suppress star formation in the host
galaxies. Any {\em observed} change in the mean SFRs due to
  luminous AGN activity will depend on the relative timescales of the
  process of shutting down star formation and of the luminous AGN activity
  itself. For example, if the time between the onset of luminous AGN activity and
  the shut down of star formation is longer than the AGN lifetime (which may in itself be highly variable; e.g.,
  \citealt{Novak11}) then it would be challenging to observe significantly
  reduced mean SFRs, simply on the basis of an X-ray luminosity
  threshold. A major uncertainty in providing a detailed
  assessment of possible connections between luminous AGNs and
  the suppression of star formation is due to the relative lifetimes of these processes. Modeling suggests
that episodes of the most luminous AGN activity last $<$80\,Myrs (\citealt{Hopkins05};
\citealt{Novak11}) while studies based on black hole mass
functions and AGN luminosity functions imply that luminous AGN
activity could last in excess of $\approx$10$^{8}$ years
(e.g. \citealt{Marconi04}; \citealt{Kelly10}; \citealt{Cao10}). Observations of AGN-driven
$\approx$1000\,km\,s$^{-1}$ outflows, found over kilo-parsec scales
in $z\approx$\,1--3 ultra-luminous infrared galaxies (ULIRGs;
\citealt{Harrison12}) indicate that star-formation episodes could be
rapidly shut down; indeed, detailed studies of local ULIRGs
with AGN-driven molecular outflows indicate that their gas reservoirs could be
completely removed in $\approx$1--40\,Myrs (e.g.,
\citealt{Feruglio10}; \citealt{Sturm11}). Therefore, while to first
order our results may appear to suggest that luminous AGNs do not
suppress star formation, it is also possible that the timescales for luminous AGN
activity and the suppression of star-formation are comparable,
meaning that any signatures of suppression would be very challenging
to detect with the current data. Clearer signatures of suppressed star formation may be more apparent when considering 
subsets of the AGN population which may represent specific
evolutionary stages (e.g.,\ \citealt{Page04, Stevens05}).

To summarize, we find no evidence for the significantly reduced mean SFRs among high luminosity 
AGNs, and the mean SFRs measured are
consistent with that expected from typical star-forming galaxies at
the average redshifts and typical stellar masses of our bins. We have suggested that
depending on the timescales involved, it could be challenging to see the
signature of suppressed star formation simply on the basis of an X-ray
luminosity threshold.
%%%%%%%%%%%%%%%%%%%%%%%%%%%%%%%%%%%%%%%%%%%%%%%%%%%%%%%%%%%%%%%%%%%%%%%%%%%%% 

%%%%%%%%%%%%%%%%%%%%%%%%%%%%%%%%%%%%%%%%%%%%%%%%%%%%%%%%%%%%%%%%%%%%%%%%%%%%% 
 
\vspace{3mm} 
\noindent We acknowledge the STFC (CMH; DMA; ADM and AMS), the Leverhulme
Trust (JRM) and NASA (MD and JK). We thank the anonymous referee, Douglas
Scott and Mat Page for useful comments. This research made use of data
from the HerMES project (\citealt{Oliver12}) which utilizes the {\it Herschel} ESA space
observatory.
 
%%%%%%%%%%%%%%%%%%%%%%%%%%%%%%%%%%%%%%%%%%%%%%%%%%%%%%%%%%%%%%%%%%%%%%%%%%%%% 

%\bibliography{final_bib.bib}
%\bibliographystyle{emulateapj}

%%%%%%%%%%%%%%%%%%%%%%%%%%%%%%%%%%%%%%%%%%%%%%%%%%%%%%%%%%%%%%%%%%%%%%%%%%%%% 

\end{document}